\pdfoutput=1 
\RequirePackage{ifpdf}
\documentclass{JINST}

\title{The PoGOLite control system and software}

\author{M. S. Jackson$^a$\thanks{Corresponding author.}\\
\llap{$^a$}Department of Physics, KTH, The Royal Institute of Technology \& \\
The Oskar Klein Centre for Cosmoparticle Physics\\
  Stockholm, Sweden\\

E-mail: \email{miranda@particle.kth.se}}

\abstract{The autonomous control system of PoGOLite is presented. PoGOLite is a balloon borne X-ray polarimeter designed to observe point sources. To obtain scientific data with optimal efficiency, independent of the ground connection, the payload control system has been made autonomous in most functions. The overall system architecture and the interconnections between components, as well as the automation philosophy and software, are described. Results of performance tests are given.}


\begin{document}

\section{Introduction}
\label{intro}
PoGOLite is a balloon-borne X-ray polarimeter tuned for measurements of point sources. The polarimeter hardware and function are described in great detail in \cite{kamae08}. The PoGOLite pathfinder is a smaller version of the originally designed instrument, and has 61 phoswich detector cells (PDCs) surrounded by 30 side anticoincidence system (SAS) units. The purpose of the pathfinder is to test the design and components, and in the future, a larger instrument with a greater collecting area may be built.  X-ray polarimetry can be used as a probe of the geometry of objects with a high magnetic field, such as pulsars and pulsar wind nebulae, and of those with an accretion disk, such as stellar-mass black hole candidates. The primary targets of the PoGOLite pathfinder are the Crab pulsar and nebula and the black hole candidate Cygnus X-1, which are among the brightest X-ray objects in the sky.

The PoGOLite pathfinder instrument was first scheduled to fly on a few-hour stratospheric turnaround flight from the Esrange facility in northern Sweden in summer 2010, but a mishap in Australia caused all NASA balloon launches to be cancelled for that year. The instrument was launched for a five-day flight from Esrange to northern Canada on 2011 July 7, but a balloon malfunction prevented the payload from reaching a sufficient altitude, and the flight was terminated after a few hours. In 2012, a similar flight was planned, but the wind at the launch facility at Esrange prevented a launch attempt. The next flight window is in 2013 July, and it is hoped that the instrument will finally complete its maiden flight.

The nature of the polarimetric measurement requires accurate pointing and a long integration time. A pointing accuracy within 0.3$^{\circ}$ is required to ensure the best possible minimum detectable polarization (MDP) \cite{cmb10}, and the the ability to distinguish source emission from the background improves with the length of the observation time. Therefore, because the balloon flight is of a limited length, and each source is observable for only a few hours each day, it is necessary to maximize the time spent observing each target and its associated background regions. 

At the beginning of the flight, it is expected that fast line-of-sight Elink communications \cite{jon09} will be possible for a limited time, ranging from a few hours up to more than a day, the longer duration made possible with a connection to the And{\o}ya rocket range in Norway. When line-of-sight communications is no longer possible, a slower Iridium RUDICS \cite{RUDICS} connection will be used. While it is expected to be possible to connect to the instrument during most of the flight, the connection could be slow or difficult to establish at times, and in order to lose as little time as possible, a control system that independently performs all functions necessary for acquisition of high quality polarization data is desired.

Control of the pointing and of polarimeter functions, and integration of all relevant systems, are provided by programs and scripts; these are installed on small Linux-based computers and make the system reasonably autonomous under all foreseen conditions.


This paper describes the software that makes the PoGOLite payload autonomous, and the specific, and, in some cases, specially made, computer and electronic hardware that runs and interacts with the software. The architecture linking the systems is of relevance, and that is described in detail. Note that functions involved with balloon- and parachute-related procedures, releasing ballast, and other functions related to piloting the balloon are controlled by the launch facility and are not discussed in this paper. This paper describes functions, algorithms, and procedures related to pointing the instrument and obtaining scientific data.


\section{Payload systems}
The three main payload systems are the payload control system (PCS), the attitude control system (ACS), and the polarimeter. The payload control system links the other systems in order to 
form an instrument capable of performing coherent simultaneous functions for the purpose of obtaining high quality scientific data. The hardware contained in the three systems and the control hierarchy under which they function is described below. 
\subsection{Payload control system}
The main software of interest in this paper is installed on devices in the PCS, which includes a credit card-sized computer called a main processing board (MPB) and two boards of the same size known as interface utility boards (IUB), as well as two Linux-based PC104 \cite{PC104} computers, each connected to an Iridium modem \cite{ir}.  The MPB and IUBs were provided by DST Control \cite{DST} in Link\"{o}ping, Sweden. The IUBs provide multiple channels of analog and digital input and output to the MPB. The MPB in the PCS is also called the payload control unit (PCU) and is described in \S~\ref{MPB}, and the PCS devices and other boards, power converters, ethernet switches, etc., are contained within a pressure vessel known as the ``PCU box". 

\subsubsection{PC104 computers}
All onboard PC104 computers contain nearly identical software, including an additional PC104 that is located outside the PCU box and is used for redundant storage. This computer is contained in a pressure vessel called the ``black box", which is attached securely to the gondola floor and is spatially separate from the PCU box. For additional redundancy, all three PC104 computers have RAID arrays comprising four solid state disks each. During normal operation, one of the PC104 computers in the PCU box functions as the primary computer, and the other is used for archiving data. In flight, the second PC104 may be used for additional functions, such as preprocessing the polarimeter data and providing the Iridium link, if the primary computer is deemed by the ground user to be overburdened. The PC104 in the black box can be used for these functions if circumstances require, but that computer is not connected to an Iridium modem, so the Iridium connections in the ACS will need to be used if the two other PC104 computers have failed. This is, however, not an expected configuration for flight. The idea behind the black box is to provide an additional redundancy for data storage, in case damage is sustained upon parachute opening or landing at the end of the flight, since the scientific data cannot be transferred to the ground during the flight.

\subsubsection{Payload control unit}\label{MPB}
The PCU is signal-based and contains a real-time operating system written by the author, which has been compiled with Borland C++ 5.0.1. It contains a 100 MHz AMD Elan SC520 processor. To it are attached two IUBs via the high-speed RS422 connection: one in the PCU box, and the other among the polarimeter electronics in the main pressure vessel. The purpose of the IUBs is to provide an interface between the MPB and various sensors, and to control auxiliary electronics boards containing power switches and multiplexed sensors. The PCU is equipped with a high-speed RS422 interface to the ACU (see below), so that signals may be shared. 

\subsection{Attitude control system}
\begin{figure}
\centering
\includegraphics[width=0.8\linewidth]{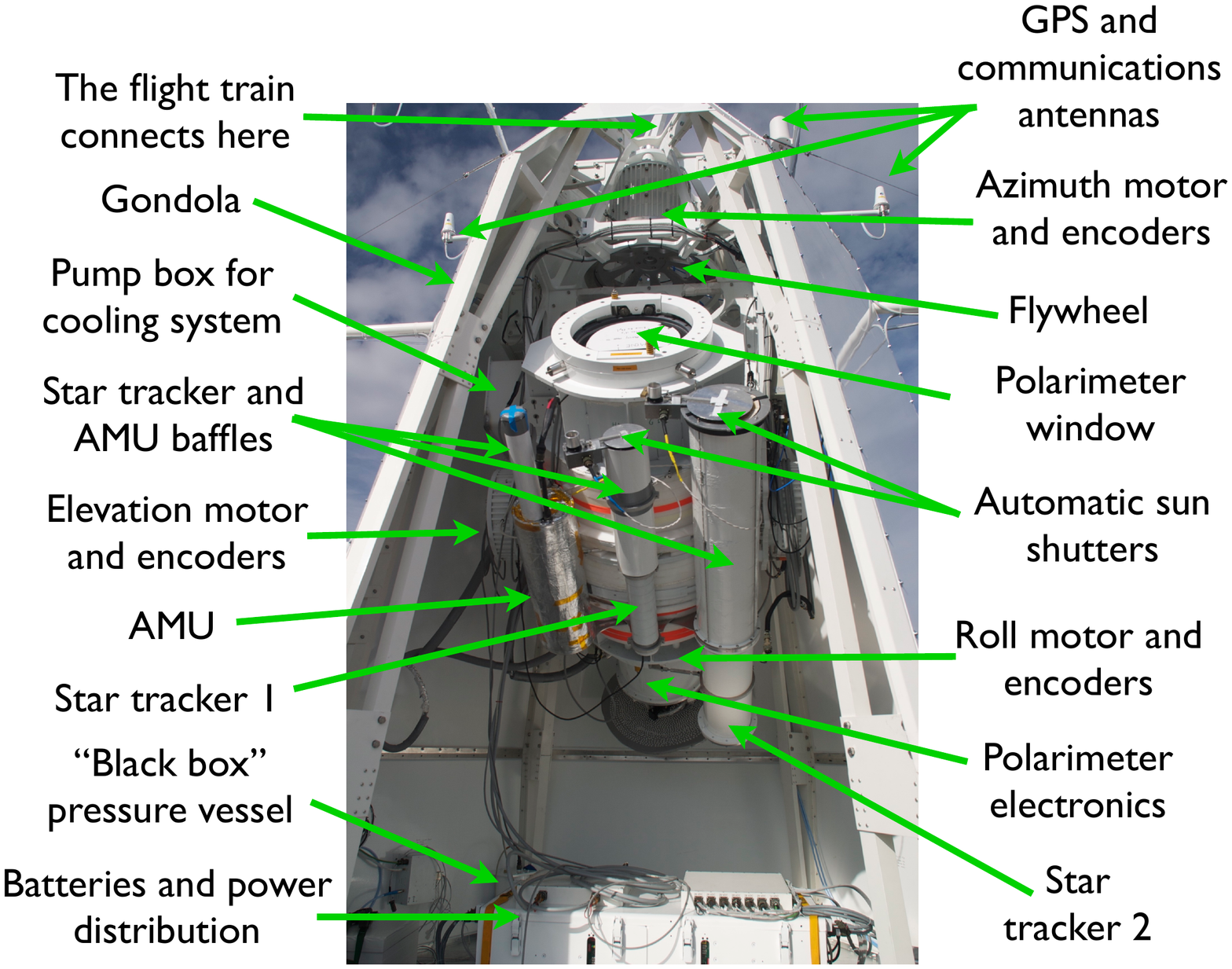}
\caption{Photograph of the polarimeter and ACS inside the gondola. The relevant visible components are labelled. The PCU box (not visible in the photograph) is attached to the top of the polarimeter pressure vessel as a counterweight. Some of the cooling fluid tubes are visible at the left inside the gondola. These transport cooling fluid out to the radiator and back to the pump. The antennas used for the differential GPS are at the ends of the long booms that are partially visible at either side of the photograph.\label{ACSfig}}
\end{figure}

The payload is equipped with a gimbal-type pointing system provided by DST, shown in figure~\ref{ACSfig}. This system, known as the ACS, makes use of motors and actuators for pointing, as well as gyroscopes, magnetometers, and a differential global positioning system (GPS) for attitude feedback. The ACS is enhanced with star trackers, and provides an extremely reliable, accurate, and stable pointing solution, and is described in more detail in\cite{strom11}. The ACS contains an MPB nearly identical to the PCU, but with different programming. This is called the attitude control unit (ACU). To it are connected multiple IUBs for controlling the motors and monitoring the pointing and health status (temperatures, currents, etc.) of the components.

\subsection{Polarimeter}\label{pol}
As previously mentioned, the polarimeter has been described in detail elsewhere. The PCU interacts with the polarimeter by means of an ethernet-to-SpaceWire interface device. All electronics boards in the polarimeter are connected via SpaceWire, and the power to the individual electronic components is controlled with the auxiliary boards described in \S\ref{MPB}. The electronics boards in the polarimeter comprise one data input output (DIO) board and twelve flash-ADC (FADC) boards.

\subsection{Control hierarchy}
\begin{figure}
\centering
\includegraphics[width=0.6\linewidth]{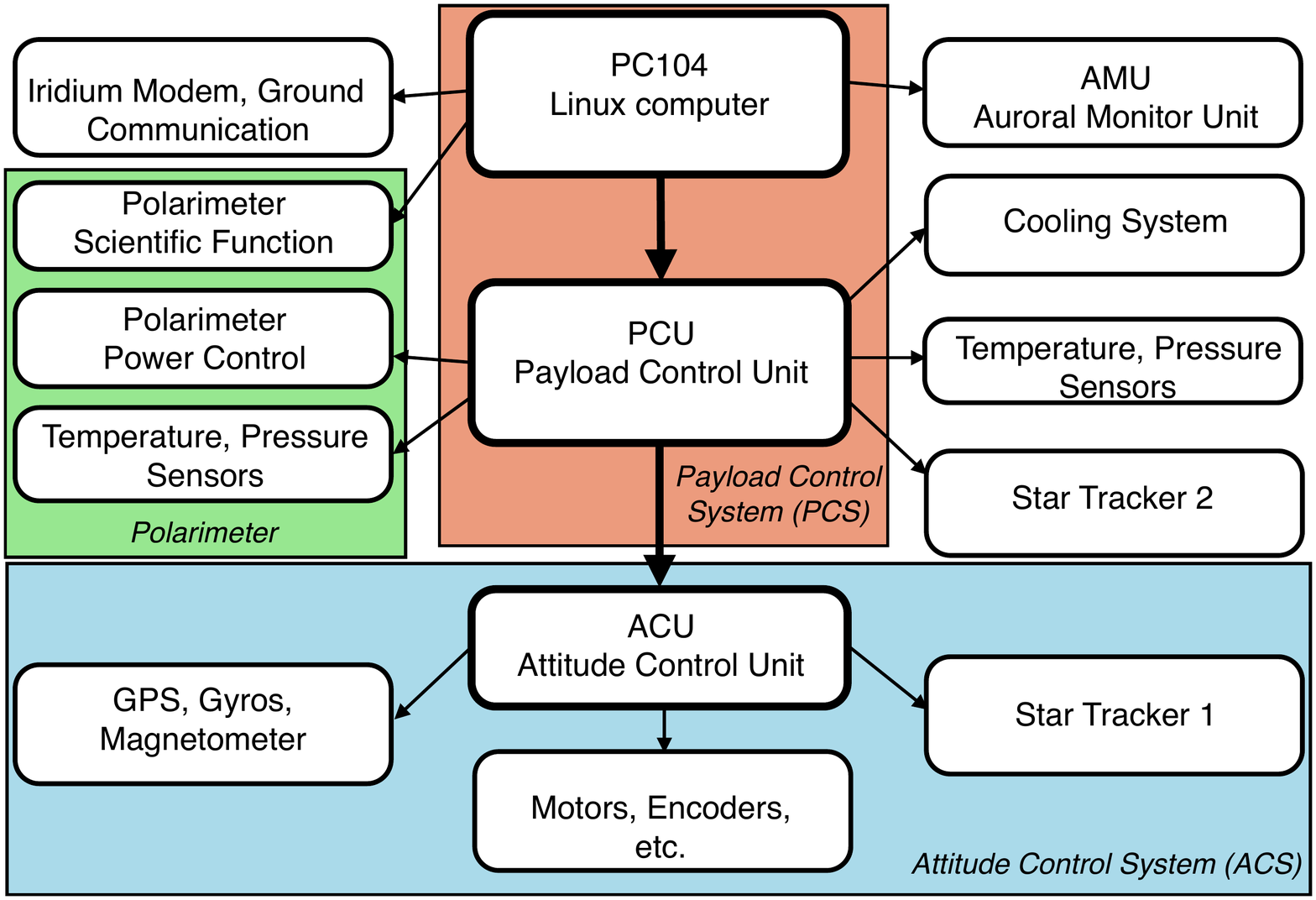}
\caption{Control structure for the PoGOLite payload. The arrow from one device to the other indicates that the device at the point of the arrow is controlled or monitored by the other device. The main pointing hierarchy is denoted by the thicker outlines and arrows down the middle of the figure. The connections indicated by the arrows are achieved through ethernet, RS422, analog voltage supply (such as for controlling the pump speed in the cooling system), or a combination of these. The PCU switches the power for most devices in the system, except those in the ACS, for which the power control is handled by the ACU, and such power connections are, for the most part, not shown in the diagram.\label{fig:hierarchy}}
\end{figure}

The control hierarchy for the payload is shown in figure~\ref{fig:hierarchy}. 
The software and scripts installed on the PC104 computer represent the highest level of control, and these are what the user usually interacts with.
The software on the PC104 and the PCU must link the function of the ACS with the polarimeter function, as well as other auxiliary but important systems such as the cooling system and the power control system. 
 As is shown in the figure, the PCU functions as the master in the high speed connection with the ACU. This means that any signals not forced by hand by an external program (not expected during flight) are set on the ACU by the PCU. This makes it possible to use a single program on a PC104 computer to manipulate signals on the PCU, which will also change the input signals on the ACU, providing full control of the attitude control system. Selected ACS output signals concerning attitude and ACS health are also shared with the PCU, and these can be accessed by the software on the PC104.

The star tracking system, described further below, consists of two star trackers pointing parallel to the axis of the instrument, each with its own camera and computer. The primary star tracker (Star Tracker 1 on figure~\ref{fig:hierarchy}) is connected through a serial link to the ACU, on which a driver processes the feedback signals into directives the pointing hardware should follow. A similar driver exists on the PCU, to which the secondary star tracker (Star Tracker 2) is attached. The processed feedback from the secondary tracker is shared as signals via the fast serial connection from the PCU to the ACU. If the RS422 link to either star tracker fails, it is possible to perform the same function with ethernet packets, though this slows down other communication in the system.

The auroral monitor unit (AMU) \cite{jok09} is shown in Figure~\ref{fig:hierarchy} for completeness. While the PC104 software controls the power state of the AMU through the PCU, and the AMU sends data to the PC104 for the purposes of monitoring, the AMU is a completely independent scientific experiment. Other than initially setting the AMU to start its own data acquisition when the balloon has reached float and automatically after any power outage, the PCS interacts very little with the AMU. The AMU is directed along the same line of sight as the polarimeter and is designed to measure the light produced by the interaction of the solar wind with the magnetic field of the earth. As well as producing scientific results in its own right, the AMU could provide a useful gauge of the background for the polarimeter. The aurora produces polarized X-rays which could give rise to a false polarimeter signal if these X-rays are directed into the polarimeter along the line of sight.

\section{Software}

The philosophy behind the software and the basic operating modes for the polarimeter and pointing system have been described in \cite{jac11}. The polarimeter modes include: 
\begin{itemize}
\item Power Save/Startup -- All polarimeter electronics are switched off to save power. This is the situation at startup.
\item Initialize -- The polarimeter electronics boards are powered on and then checked for proper functionality and power cycled if necessary, until all are on and ready.
\item Ready -- The polarimeter is powered on and ready to take data.
\item Acquisition -- The polarimeter is actively acquiring data.
\end{itemize}
The reason for defining discrete modes is that the ACS is signal-based, and therefore it is simpler to think of an ACS function in terms of a named mode that corresponds to an integer. Since there are so many functions involved in the polarimeter operation associated with each overall mode, controlling the system, including the polarimeter, by means of individual commands is more useful, though the predefined command lists in the scripts (see below) may be grouped into procedures that could be thought of as belonging to different modes. These commands include changing and monitoring the ACS mode, which comprise the following:
\begin{enumerate}
\addtocounter{enumi}{-1}
\item Startup -- The default mode upon startup.
\item Initialize -- The ACS must be initialized before it can perform other tasks. This initializes each motor in turn by finding its reference point.
\item Stow stabilized -- The polarimeter is directed directly upwards and held in place by magnets, and the azimuth is held at a chosen angle
\item Stow unstabilized -- Same as above, but the azimuth is not specified and the payload can rotate freely.
\item Exercise -- The motors in the ACS are gently moved back and forth to prevent ice formation. This is useful primarily during the ascent when the temperatures are very cold in some layers of the atmosphere.
\item No control -- The motors are not engaged, except the roll motor which can be used if initialized.
\item Azimuth/elevation pointing -- The payload points to a specified azimuth and elevation.
\item Right ascension/declination pointing -- The ACS is given the celestial coordinates of an object, for which it calculates the azimuth and elevation and points appropriately.
\item Power save -- The power to the motors is switched off to avoid running down the batteries if the solar cells are not functioning.
\end{enumerate}

The software described in this paper consists of a main control program and script on the PC104, and the real-time program in the PCU. The ACU software was supplied with the ACS by DST, and provides an extremely stable means to control and monitor the pointing, but is described in this paper only in terms of its inputs and outputs for specific functions initiated with the PCS computers. 

\subsection{Main control program}
The main control software has been written to take simple 3-character text commands along with a number of parameters as input, either typed by a user or contained in a script. Some examples of commands include `con' to connect to the PCU, `daq' to start data acquisition, `sen' to dump all current sensor outputs to the screen, and `pnt' to point at a target. A `hel(p)' command provides a comprehensive list of commands, including all possible combinations of input parameters, and also allows the user to look up a specific command.

The program runs a number of threads for the purpose of performing multiple tasks independently, such as pointing, preparing the polarimeter components for data acquisition, checking various housekeeping values, and waiting for and processing command input. There are three main components of the software, organized into different source files. The first is the``wrapper" component, which runs the threads, processes input command scripts, and passes on the command input. The second provides a class that parses the commands and provides a link for setting and reading signals on the PCU computer via the Modbus protocol. It also calls functions from the third component, which communicates with the ethernet-to-SpaceWire interface device in the polarimeter in order to perform all the polarimeter functions. The SpaceWire/RMAP library \cite{taky10} is used to provide the SpaceWire functionality.

\subsection{PCU software}
The real-time PCU software is signal-based, and performs some functions based on the values of and changes in certain signals. Most of the actions are done by the main control program, however. The function of the PCU software is predominantly to gather sensor information (temperatures, pressures, etc.) from various sources and store the information as signals, sometimes after converting the information into a more human readable form, and to provide an interface to the ACS, power switches, and other devices, through output signals. Most of the approximately 250 input and output signals stored on the PCU are manipulated or read by the main control program, which performs most of the higher-level functions. 

\subsubsection{Timing the polarimeter data with a gated PPS signal}
The PCU receives a pulse per second (PPS) signal from the ACU. This signal arises from the GPS receiver, and the rising edge of the pulse can be used to measure each second in real (UTC) time with an accuracy of better then one microsecond. These pulses are in turn output by the IUBs connected to the PCU. One of these signals is sent as is to the AMU, while the other signal destined for the polarimeter electronics is gated (i.e. the PPS signal is blocked) by the PCU software in a particular pattern which repeats every 5 seconds. In addition, the first second of every minute is gated, and these patterns enable the absolute time to be reconstructed from the polarimeter data. Since the approximate starting time (to the nearest few seconds) of each polarimeter data run is recorded, this allows the absolute time to be reconstructible by recording this pattern in the polarimeter electronics for ten seconds before each run is started. By comparing the internal clock times of PPS and event signals recorded by the polarimeter electronics (which are stored in the data files), and considering delays introduced by the electronics itself, each event can be timed with an accuracy of significantly better than one millisecond, which is sufficient to reconstruct the light curve of the Crab pulsar.

\subsection{Control script and automated function}\label{auto}
\begin{figure}
\centering
\includegraphics[width=0.3\linewidth]{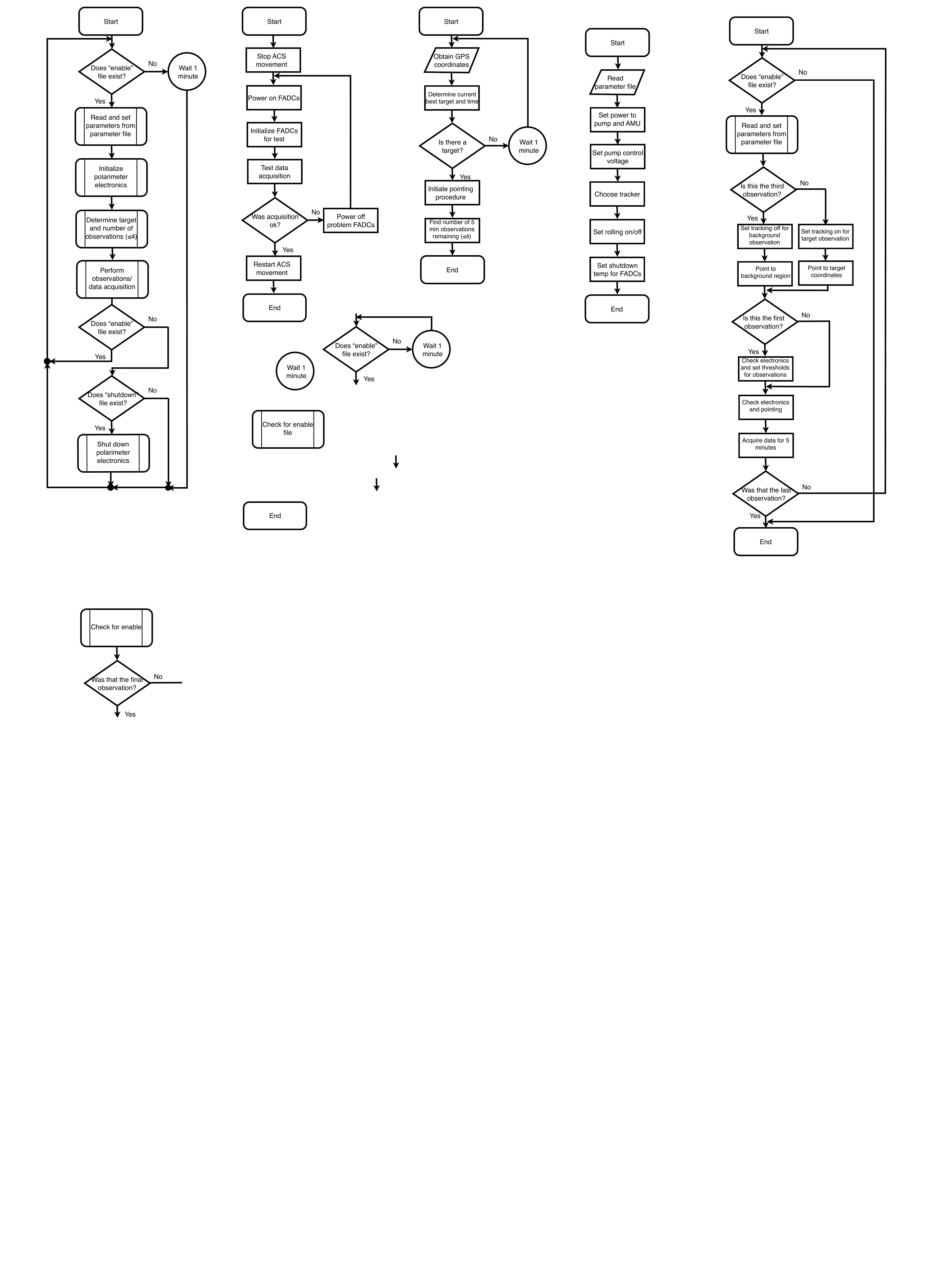}
\caption{Flowchart for the automated function.}
\label{fig:autoflow}
\end{figure}
As previously mentioned, the main control program described above takes command input typed by the user or loaded into a script file. The program continuously checks to see if a file of a specific name exists, and if it does, it copies the commands from that file into a new file used and modified solely by the program. As the commands are completed, they are removed from the file. If a command does not complete, or it crashes the program, the command remains at the top of the command list and it is tried again. For this reason, it is very important that all the commands in the script file be valid. This is accomplished by making a collection of working scripts and then copying these into the filename the program expects as the tasks are required.

The automation script is a bash script that copies premade scripts into the filename that the program reads. If a command at first fails for any reason, the script will pause until the program completes the command. Under most circumstances, the program will either have crashed or had a temporary communication problem with another device. If the former, an init script ensures that the program restarts, and it will attempt the command again. If the latter, the program tries once more, and this will usually result in a successful completion of the command. 

The algorithm followed by the automation script is indicated in figure~\ref{fig:autoflow}. The script checks for the existence of an ``enable" file when the script first starts, and before and after every observation, so that the user need only rename this file in order to pause the script. The user has the option of also creating a ``shutdown" file, which will indicate to the script to shut the electronics down when the script pauses as a result of not being able to find the ``enable" file. To restore the functionality of the script, the ``enable" file is again placed in the appropriate directory. The existence of the ``enable" and ``shutdown"  files can be controlled with a script, so the state of the system can be modified with a single command from the ground.

After checking for the existence of the ``enable" file, the script sets some parameters it reads from a parameter file. The parameter file is read often over the script operation, so that the user can change parameters on the fly without having to disable or restart the script. Parameters include settings of whether to power the AMU and the pump, the pump control voltage, which star tracker to use, how many FADC boards to use, and others, and the purpose of changing this file is usually for ground tests, but it can be changed during the flight if a device malfunctions.

The automation script then powers the polarimeter electronics and runs initial tests on them to ensure they are functioning correctly. Possible error states in the electronics boards include high noise levels in certain channels which can simulate detector signals, and failure to initialize data stored at particular addresses, usually caused by a failure of the programming to be copied from the ROM into active memory. The boards that are found to have not initialized correctly are power cycled and the process is repeated until all boards check out. This procedure has been found to be almost always successful, but after a few tries, any boards that are still malfunctioning are left switched off until the next cycle. Once the boards have been running, it is necessary to check them in this way only every few runs.

After the polarimeter is initialized, the script sends a command that retrieves the GPS coordinates from the main control program. The coordinates are saved in a file that the script reads and then sends as input to a second smaller program, which then uses the coordinates and the current time to choose the best target from a predefined prioritized target list. The program outputs the number of the current best target and the remaining time for which the target will be valid. The script then sends commands to the program, which will point the instrument at the target and nearby background regions in a predefined timed pattern, described in \S\ref{pointing}, and the instrument acquires polarimeter data for each pointing. After a cycle of four target and background observations, the script tests the polarimeter electronics as above and the cycle begins again. This pattern is repeated indefinitely until one of the following: 1) the user sends a command from the ground to stop the script by changing the name of the ``enable" file, 2) the program gets stuck on a particular command, because of a communication problem or other error, or 3) the power to the payload is disrupted. In the first two instances, user intervention will be required to restart the automated function, and log files are available so that the user can quickly learn of the problem. In the third instance, the system should resume function once the power has been restored.

\section{Housekeeping and system health}
To correctly process the polarimeter data and to monitor the system from the ground, it is necessary to log and monitor certain signals and parameters. This is done automatically by various components of the system, and a brief overview of the functionality is provided below.

\subsection{Logging}
Through the connections to the IUBs in the PCU box and in the main polarimeter pressure vessel, the PCU can monitor temperatures, pressures, and the cooling fluid flow rate. In addition, polarimeter data acquisition status parameters, such as run number and  event rates, and pointing status parameters, such as the position and pointing coordinates and tracking status, are saved internally in files on the PCU. Some signals that are expected to vary slowly, such as temperatures, are recorded every ten seconds, while signals necessary for the reconstruction of pointing, such as positional and directional data, are stored four times a second. These rates were chosen to provide sufficient information while not filling up the hard disks in the onboard computers. Both types of logs contain the GPS time for each entry. Once each of these files is full, it is sent via ftp to one of the onboard PC104s. The choice of which PC104 to send to is made by stepping through the servers that are powered on until the file is transferred successfully. If no servers are found, the log files are eventually overwritten on the PCU and the data are lost. Once a script on the PC104 finds a new log file as received from the PCU, it renames the log file with an appropriate date and time stamp. 

In addition to the logs mentioned above, additional log files are stored on the PC104s. These include a log of the health status of each PC104 computer, including the RAID and filesystem status, and logs written by the main program and control script described above. A script running on each PC104 ensures that all log files are compressed and stored on all three onboard PC104 computers for maximum redundancy. 

Every few minutes, whether through the Elink or Iridium connection, the new log files are automatically retrieved by a ground server. The data are graphed as they are available, allowing the ground user to visually monitor most aspects of the pointing, polarimeter function, and health status.

\subsection{Cooling system}
The polarimeter electronics boards use a significant amount of power, and therefore they generate heat. It is estimated that the boards generate a few hundred watts of heat, and since they are enclosed in a sealed space with no ventilation, it is important that they are cooled so that they and the PMTs, which also generate heat, but to a lesser extent, do not get too hot. It is estimated that the optimal operational temperature for the PMTs is around a few degrees celsius, because a low temperature minimizes the dark noise rate in the tubes. By observation, the FADC boards tend to cause error states, requiring them to be reset when they reach approximately 55$^\circ$C, as measured at a chip known to be one of the hot points on the board. In addition, some of the FADC boards do not function properly when they start below room temperature. This narrow operating temperature range arises because the boards were originally intended to be used only as test boards, and are not designed to be used in harsh environments. To keep the components at an optimal temperature, a radiative cooling system is employed. A method for providing heat is not required, because the boards produce a significant amount of heat when they are powered on, and the rate at which they dissipate heat is greater than the rate at which heat can escape through the pressure vessel walls and be carried off through convection or radiation. Any boards that malfunction from the cold are simply power cycled automatically until they begin to function properly, since the other boards that are powered up will tend to increase the temperature inside the pressure vessel after a cold start.

The cooling system pumps fluid through a radiator and through the main pressure vessel, with significant thermal contacts to the PMTs and to the polarimeter electronics boards. The radiator is made of aluminum and is approximately two meters wide, one meter long, and a centimeter thick. The radiator is intended to point away from the sun during observation of both of the main targets. It is foreseen that the cooling system will be operational during all the time while the FADC boards are powered. However, it is possible for the system to control the cooling function based on the temperature readings from the main pressure vessel. 

Because of the rarefied atmosphere at float, radiation is the only efficient means by which the instrument can expel heat, even though the temperature at float is around -20$^{\circ}$C. For this reason, it is very important that the radiator always be pointed away from the sun while the instrument and cooling system are operational. The cooling system will automatically be halted whenever the fluid in the radiator is at a higher temperature than in the pressure vessel. The purpose for this is to prevent fluid heated by the sun to be pumped through the system. 

\subsection{Health monitoring}
The temperatures are regularly monitored by both the PCU software and the main control program. If the temperatures of the FADC boards become too high, the main software automatically ramps down the PMT voltages and switches off the power to the PMTs and FADCs. If for some reason, the main software does not do this, whether it is because the software is not running or for another reason, the PCU will shut the FADC boards down at a much higher temperature that is close to the temperature at which damage may be done. Unfortunately, the PCU is not capable of gently ramping down the PMT voltages, so this is to be considered a last resort.

The PCU also has access to health information from the ACU, such as motor voltages and currents. However, since the ACU monitors its own health and temporarily shuts down any overheating or malfunctioning components, it is not necessary for the PCS computers to contain any functionality regarding ACU health other than to monitor ACU error signals and pause the operation until the errors clear.

\section{Pointing}\label{pointing}
\begin{figure}
\centering
\includegraphics[width=0.32\linewidth]{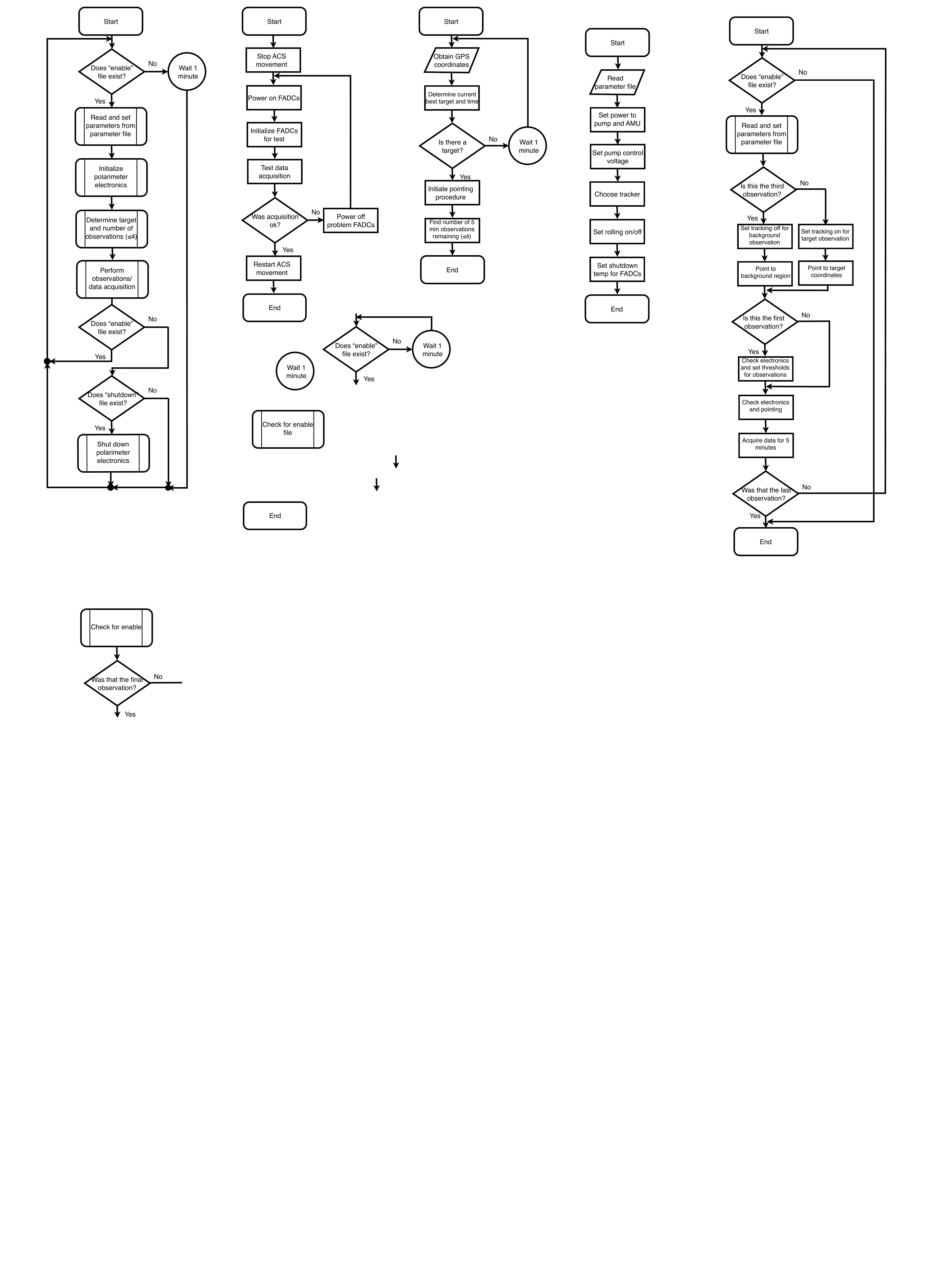}
\caption{Flowchart for the pointing and data acquisition during the automated function.}
\label{fig:pointingflow}
\end{figure}
The pointing scheme for each target is a pattern that lasts around 20 minutes, and this is repeated in a cycle while the target is valid (i.e. while the target is at a high enough elevation and while no higher priority targets rise). The pointing algorithm is illustrated in a flowchart shown in figure~\ref{fig:pointingflow}. The scheme for pointing and data acquisition normally includes four five-minute observations, of which all are performed at the coordinates of the target object except the third, which is performed with the instrument pointed at a nearby predefined background region. There are either two or four predefined background regions per target, and the algorithm steps through them with each cycle (this is not indicated on the flowcharts). The purpose of obtaining data from background regions is to provide a means of measuring the modulation curve for an unpolarized field, to be sure that it differs from data from the possibly polarized object. 

By this procedure, the target source is observed for three quarters of the time, and the background regions are observed for the remaining one quarter. This allocation of time has somewhat arbitrarily been chosen, and can be changed in flight during the line-of-sight connection phase after the first data have been retrieved and processed. A seemingly large amount of time has been allocated for background measurements because it is very important to understand the background profile at this latitude, as it has never been measured. It is expected that the background will change depending on the elevation of the source, since the elevation represents varying atmospheric mass, and the background measurements can be compared with previous observations at lower latitudes and with simulation.

The background measurement is performed during the third observation for two main reasons. The elevation, which is constantly changing, will be at a level intermediate to all source observations during the third cycle. In addition, if the observed source is in a favourable location in the sky for only ten minutes, for example, this scheme gives precedence to obtaining data from the source, since, in this case, only two observations are made, both of them on the source.

\subsection{Star tracking and tracking modes}
It is possible to point the instrument within a few tenths of a degree with the ACS, and when one of the star trackers is used, the accuracy is better than one tenth of a degree, which exceeds the requirement for the highest quality polarization data. This accuracy is sufficient to ensure the lowest possible MDP for any given target. To perform a pointing function, the ACS is given the coordinates (either celestial or azimuth and elevation) and then it is set into the appropriate pointing mode for the type of input coordinates. Celestial coordinates are automatically converted by the ACU into azimuth and elevation with the use of the GPS time. Differential GPS, gyroscopes, and magnetometers are used in that order of priority to determine the azimuthal angle. It is possible to track on stars in either of the pointing modes. The azimuth/elevation tracking was added for ground testing purposes, because many tests must be done indoors and the time of year at which the balloon is intended to be launched does not allow for tracking tests on stars since the sun never sets at the launch facility.

The star trackers can be used in a few different modes, corresponding to which trackers are being used, and how the tracking is initiated. The software used on the star tracker computers is nearly identical for both trackers, varying only in such parameters as field of view and magnification, because different lenses are used on the two tracker cameras. The default mode for user-controlled pointing is that the tracking is initiated by the user when the star field is observed to match the field corresponding to the desired pointing coordinates. Once tracking is started, the ACS responds to coordinates provided to it by the trackers as feedback for the pointing quality, and ignores all other input from the differential GPS, etc. If the tracked source moves too far away from the pixel in the tracker camera that corresponds to the correct sky coordinates, the tracking process is halted and the ACS reverts to its normal input scheme for the pointing.

The star tracking system also has an automatic mode, and this is used in conjunction with the automated function. In automatic mode, the ACS sends a signal as an ethernet packet to the active tracker when the pointing is determined to be accurate and stable, and the tracker begins to send feedback coordinate data for the source initially closest to the predefined coordinates on the tracker camera.

\subsection{Sun avoidance}
Other than to observe the sun for scientific reasons (see below), it is desired that pointing directly at the sun generally be avoided. One reason for this is that the AMU contains a PMT and is particularly susceptible to solar exposure. For this reason, a command is automatically sent through a serial connection by the main program to the AMU to switch off the PMT, should the instrument be pointed within 15$^\circ$ of the sun. When the danger has passed, another command is sent to switch the PMT back on. Since the instrument moves quite slowly azimuthally, a transit of the sun while pointing to a new target could produce undesired thermal effects in the polarimeter.

So that the AMU does not need to switch its PMT on and off, and to avoid any thermal effects, a simple sun avoidance algorithm has been implemented for pointing toward a new target. Since the flight latitude is so great, the sun will be quite low on the sky at all times, so the instrument is first pointed toward an elevation of 60$^\circ$ at its present azimuth. The gondola is then rotated to the new azimuth while maintaining the elevation, and then the elevation is set to that of the new target, and ordinary pointing using celestial coordinates is initiated. If the new target is within 15$^\circ$ of the current pointing location, the instrument is pointed straight to the new target without first adjusting the elevation.

If a target with coordinates within 15$^\circ$ of the sun is input, the program will reject the pointing command. The targets in the list with which the automated script is provided have been thoroughly planned and checked, and none lies too close to the sun at the time of year of the planned flight. For safety, the instrument can only be pointed directly at the sun with a specific command, as described in the next section.

\subsection{Sun pointing}
If a solar flare or other phenomenon should occur, it is possible to point the instrument at the sun with a special command. While this is unlikely to put the polarimeter or star trackers at risk because of the shielding on the polarimeter and automatic sun shields on the trackers, it is possible that prolonged sun observation could cause undesired heating, particularly at the front of the polarimeter. Since the main scientific target is the Crab, which is quite close to the sun (within 20$^\circ$) at the time of year of the flight, precautions have already been taken to protect the components from sun exposure. However, it is foreseen that a sun observation would only be made if it is deemed to be scientifically interesting, and only after a sufficient amount of polarimeter data has been obtained from the primary targets to be worth the risk. Because the temperatures and other parameters must be carefully monitored while observing the sun, sun pointing is not part of the automated function and must be initiated manually from the ground.

\subsection{Pointing system, tracking, and automation tests and results}

Because the automated function requires that the pointing system and tracking work correctly, it is not meaningful to test the automation without the other two elements. However, the ACS can be tested and tuned independently. The addition of star tracking to the pointing solution adds an additional layer of complexity, and the tracking, particularly in the automatically initiated mode, is required to be thoroughly tested along with the pointing before the automation can be tested.

Pre-launch tests of the pointing system with the star trackers is problematic, as mentioned before, because the launch facility is above the arctic circle and the sun never sets in the summer. The tracking system was tested in Link{\"o}ping during the winter in early 2012, but some bugs and other difficulties prevented complete end to end tests of the entire fully functioning system. For this reason, it was necessary to do tracking tests within the month before the planned launch by pointing at the sun. The star trackers are equipped with sun shields that activate automatically when the trackers point too close to the sun and an independent light meter measures a high signal. The shields are each equipped with a small aperture over which is installed a mylar filter. This allows the system to track the sun. A few difficulties result from this method. The first is that the sun has a large angular extent that is not accounted for in the tracking software, which is designed to lock onto point sources. Another minor difficulty is that the celestial coordinates for the sun continually change, but this effect is significant only over a number of hours, so that tracking tests that last for less than one hour are possible without redefining the coordinates. The tracking system was found to behave basically as expected during the test. Another positive result of this test was that the system did not experience a significant increase in temperature in any of the components. However, this is not necessarily an indication that the situation will be the same when the instrument is in the stratosphere.

Another testing method of the tracking was employed indoors. Desk lamps were used as "stars", and the timing of the system was frozen in order to accommodate the fact that the lamps do not move on the sky. The coordinates also needed to be adjusted, because the differential GPS and magnetometers do not function properly indoors. However, even with the difficulties, it was possible to move the pointing from one ``target" to another and back with good accuracy. The autonomous function was tested with the lamps as listed targets, and the tracking and data acquisition systems functioned acceptably over the duration of the tests, which were a few hours at a time.

It is foreseen that further tracking and full automation tests will be done using stars in the night sky in early 2013, while the nights are still long at the launch facility. The amount of testing time this affords should be sufficient to make a measurement of the pointing accuracy and behaviour under various circumstances. One limitation is that the gondola must be hung outside in order for the ACS to be able to use the differential GPS and the magnetometers, which undergo interference indoors. A wooden hang frame has been produced to support the gondola and accommodate the booms on either side, without producing interference by means of reflections as a metal frame would. The frame has been kindly built and provided by Esrange personnel. The gondola, with the solar panels installed vertically at the sides, is particularly susceptible to changes in wind speed and direction, so it can only be used outdoors in very calm conditions. Another limitation of pointing tests done at night is that the gondola will need to use only battery power, since the solar panels will not provide power without the sun. This is not an issue during flight, since the sun will always be in the sky.

\section{Discussion}
This paper describes a robust, autonomous control system for PoGOLite. The purpose of the autonomy is to ensure that scientific data are obtained for the greatest possible portion of the flight time, and to ensure that all components are kept within their optimal operational conditions, independent of connection and other factors. It also means that users not familiar with the system can obtain scientific data and need only monitor the available information sent to the ground, rather than make decisions about which target to observe and learn and reproduce a sequence of commands. The autonomy is made possible by the robustness of each system against irrecoverable errors and failure. From the test results, it can be expected that the PoGOLite instrument will perform as designed, including the automation, and even if the communications fail for much of the flight, it is likely that good quality scientific data will be stored onboard and recovered at the landing site.

\acknowledgments
I would like to acknowledge the hard work of Stefan Rydstr{\" o}m, who designed and programmed the auxiliary electronics boards mentioned in \S\ref{MPB}, designed the hang frame, and provided invaluable support (of a technical nature and otherwise) throughout the duration of this work. The personnel at DST Control have worked countless hours to ensure that the ACS and the interface between the ACS and PCS work flawlessly, especially Jonas Lind, who wrote the ACU software and provided guidance for writing the PCU software; David Steen, who provided the high-speed interface between the ACU and PCU; Olle Wellin, who tuned the ACS; and Jan-Erik Str{\"o}mberg, who provided the ACS contract, design, and hardware. G{\"o}ran Olafsson and Hans-Gustav Flor{\'e}n at Stockholm University provided the star trackers and tracking software, as well as information about guide stars. The polarimeter DIO and FADC boards mentioned in \S\ref{pol} were programmed in VHDL by Hiromitsu Takahashi from Hiroshima University. Some of the SpaceWire functionality was taken from test code originally written by Takayuki Yuasa. This work has been made possible by a grant from the Swedish National Space Board, and work on the ACS was funded by the Swedish Research Council.





\end{document}